# Faraday-Ramsey rotation measurement in a thin cell as an analogy to an atomic beam

**MARK DIKOPOLTSEV,[1,2,†] ELIRAN TALKER,[1,†], YEFIM BARASH[1], NOA MAZURSKI[1] AND URIEL LEVY[1,*]**

[1]*Institute of Applied Physics, The Faculty of Science, The Center for Nanoscience and Nanotechnology, The Hebrew University of Jerusalem, Jerusalem 91904, Israel*
[2]*Rafael Ltd., IL-31021, Haifa, Israel*
[†]*These authors contributed equally.*
**ulevy@mail.huji.ac.il*

**Abstract:** Atomic beams are powerful tools for measuring spin coherence in hot vapors but require bulky setups, limiting device miniaturization. We demonstrate that micron-thin vapor cells can mimic atomic beam behavior by exploiting geometry-dependent velocity filtering. In a 5 μm rubidium cell, coherence is preserved for atoms moving parallel to the cell walls, enabling observation of the Faraday-Ramsey effect without buffer gas or anti-relaxation coatings. Using a spatially displaced pump-probe scheme and magnetic field scanning, we achieve clear Ramsey fringes and validate our model experimentally. This technique offers a compact alternative to atomic beam systems, supporting scalable sensors and frequency standards.

## 1. Introduction

Atomic beams, consisting of alkali atoms, have a wide range of applications in modern scientific research and technology, including studies of parity violation [1,2], high-precision measurements of fundamental constants [3–5], quantum information [6], and atomic clocks [7–9]. These systems are effective due to their long coherence times and spatial selectivity. However, they often require bulky setups involving ultra-high vacuum systems and beam collimation elements. This complexity limits their practicality for compact or portable devices.

In parallel, optically transparent, hot alkali vapor cells are also widely used in a variety of atom-photon interaction applications. These include portable atomic clocks [10–12], miniaturized magnetometers [13,14], and quantum gyroscopes [15–17]. Advances in MEMS technology have enabled the fabrication of such cells, offering simplicity, reliability, and versatility. The vapor density in such a cell is typically controlled by adjusting its temperature. [18].

The geometric configuration of the cells significantly impacts the lifetime of atomic spins. Factors such as the shape and size of the cell, its optical depth, and atoms spin destruction due to the collision with the cell walls, influence the longevity of spin lifetime. These limitations restrict further miniaturization. Extensive theoretical and experimental research has been conducted over the past decades to explore the possibilities for miniaturizing hot alkali vapor applications [19]. However, in miniaturized cells, frequent atom-wall collisions often pose a significant obstacle to the development of high-sensitivity sensors.

The ability to fabricate ultrathin alkali vapor cells, on the order of microns [20], endows unique features, making them valuable tools in specific research fields and applications. These cells enable the exploration of high alkali atom densities for studying all-optical switching and isolation [21] as well as nonlinear frequency conversion [22]. The thin cell geometry allows for 2D millimetric-scale magnetic gradiometry, which is usually achieved with separate cells [23]. Additionally, these cells provide a controlled environment for studying the properties and interactions of Rydberg atoms [24]. Atomic beams can also be generated using spatial filters, getters, and separate cell regions, [25]. The thin-cell geometry enables the selection of velocity directions, facilitating sub-Doppler spectroscopy [26,27] and the narrowing of hyperfine and Zeeman transition resonances [28,29].

The high rate of spin destruction due to wall collisions necessitates in-depth investigations into decoherence rates. These rates depend on factors such as wall composition and the presence of anti-relaxation coatings. To mitigate the effects of wall collisions destruction, spin coherence can be preserved using buffer gas. This changes the atomic motion from ballistic to more diffusive behavior and potentially increases the mean time for wall collisions. However, the presence of buffer gas atoms can affect long-term stability due to temperature-dependent shifts in hyperfine atomic levels [30,31]. Additionally, buffer gas broadens the optical transition linewidth and mixes optical interaction with different hyperfine levels, making this technique unsuitable for applications that rely on interactions with single hyperfine levels or depend on levels separation [32,33]. These limitations motivate alternative methods for preserving coherence, such as anti-relaxation coatings or geometrical approaches like velocity filtering. Approach, involves applying anti-relaxation coatings, typically made of paraffin-like materials, to the cell walls to reduce the rate of spin destruction due to wall collisions [34–36]. However, this method is constrained by fabrication limitations, specifically the low temperatures that typical coatings can withstand during the fabrication process. Further reduction in cell dimensions leads to a dominant wall collision rate, which restricts spin lifetime.

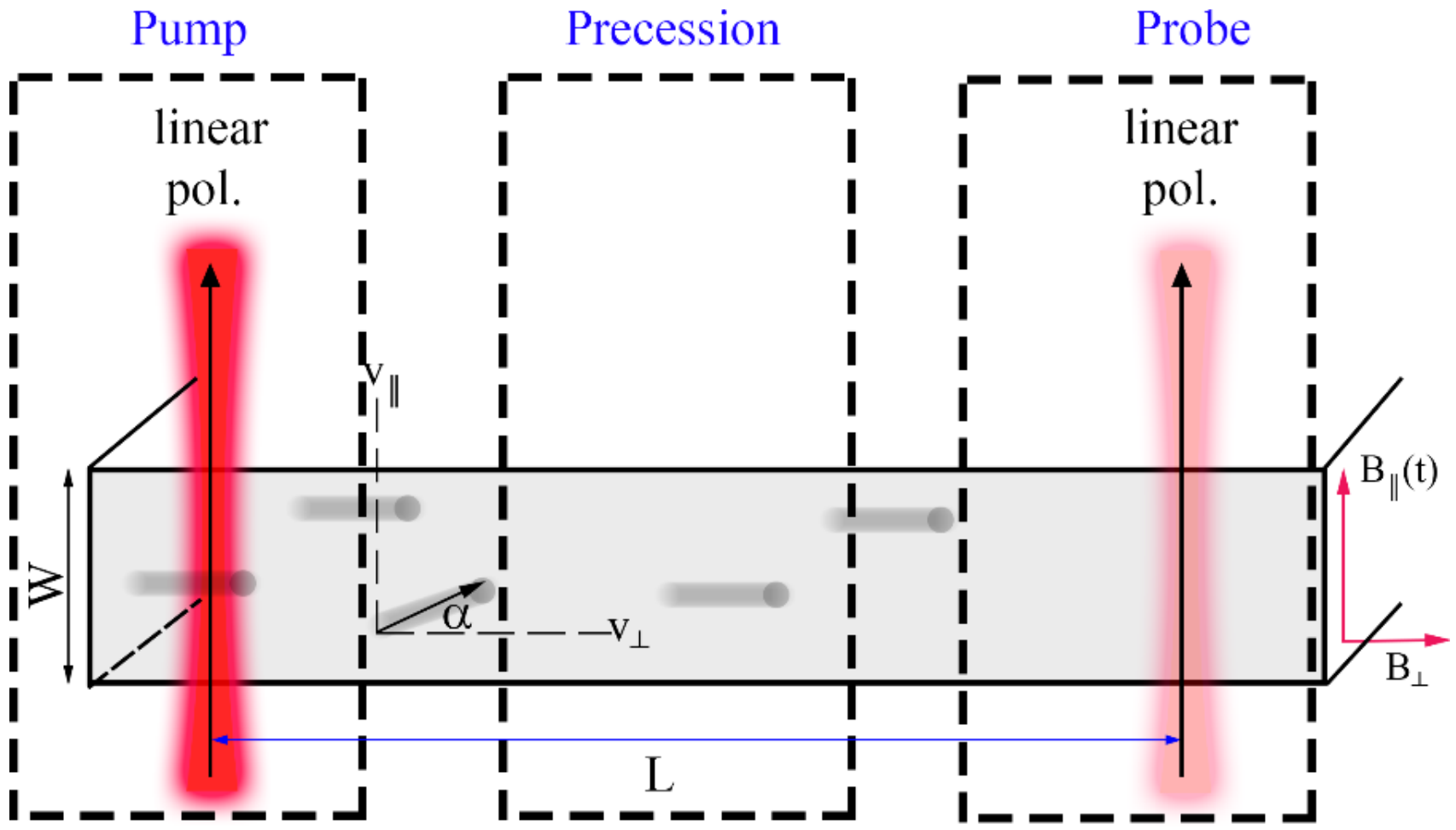

Fig. 1 Cartoon illustration showing the working principle of Faraday-Ramsey spectroscopy in a thin cell with width $W$ and distance $L$ between the pumping and probing areas. In the optically pumped volume, atoms achieve a specific distribution of Zeeman states. In the precession area, only atoms with velocities perpendicular to the beam propagation direction (defined as $v_\perp$) and parallel to the cell's long axis, propagate directly towards the probing beam. These atoms experience circulation of their Zeeman level populations under the influence of the total magnetic field $B = \left(B_\parallel^2 + B_\perp^2\right)^{1/2}$. The magnetic field component parallel to the beam direction, $B_\parallel$, is slowly varied to scan the NLFE effect, while the perpendicular component, $B_\perp$, remains constant. The symbol $\alpha$ represents velocity direction. Atoms that reach the volume illuminated by a probing beam undergo Faraday rotation, which depends on the Bloch rotation of their Zeeman levels.

Contrary to efforts aimed at minimizing the destructive effects of wall collision, this study leverages them as spatial filters for atomic coherence through spatial velocity filtering. The thin cell has a width $W$, which defines the transverse confinement of atoms, and a length $L$, which represents the separation between the pump and probe beams. Atomic motion in the thin cell is characterized by two velocity components: $v_\parallel$, the velocity parallel to the propagation direction of the pumping beam, and $v_\perp$, the velocity perpendicular to it. Since coherence is preserved primarily for atoms moving parallel to the cell walls, filtering atoms by their velocity components effectively emulates an atomic beam. Thus, the thin cell selectively transmits atoms with well-defined velocities, mimicking the collimation properties of atomic beams without a physical aperture or a vacuum. $W$ and $L$ together determine velocity filtering and coherence time. This technique enables observation of Ramsey fringe patterns, which are typically measured by magnetic field scanning in atomic beams or centimeter-sized alkali cells – a phenomenon known as the This Faraday-Ramsey effect.

## 2. NLFE Theory

Nonlinear Faraday effect (NLFE) experiments rely on the high sensitivity of atomic spin coherence measurements, making them powerful methods for high-resolution studies [37,38]. The NLFE effect, as described in [39], involves three key steps, illustrated schematically in Fig. 1. The first step, labeled the ‘Pump area’, generates spin coherence through optical pumping with linearly polarized resonance radiation. As shown in Fig. 2, this process aligns atomic populations, specifically creating $\Delta \mathrm{m} = \pm 2$ coherence in the ground state of a rubidium atom. The second step, referred to as ‘Precession’ in Fig. 1, describes the evolution of this coherence as alkali spins interact with an external magnetic field during atoms propagation through the cell. This interaction induces spin precession at the Larmor frequency, which is central to the NLFE response. The third and final step, labeled ‘Probing’, involves optical detection of the medium’s altered properties. In this spatially separated probe region, the atomic alignment is measured using a second laser beam and a polarimetric setup based on the Faraday effect. This approach, known as Faraday-Ramsey spectroscopy, enables detection of time-averaged alignment. The ensemble of aligned atoms acts as a medium with polarization-selective properties such as birefringence and dichroism, which facilitate optical readout.

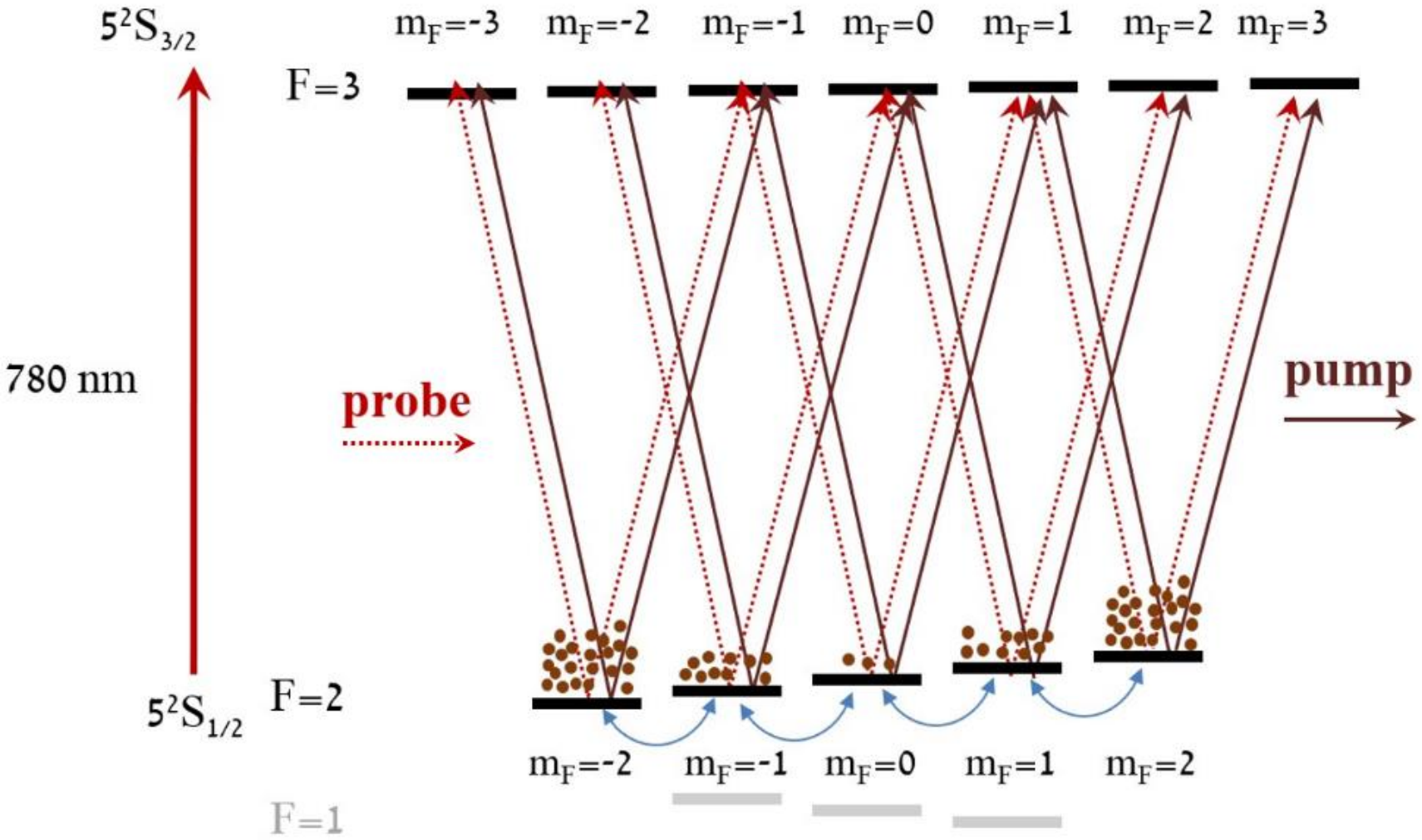

Fig. 2 Transition of $^{87}$Rb on the D2 line from $F = 2$ to $F' = 3$, illustrated with Zeeman splitting. A detuned, linearly polarized pump beam, indicated by brown solid arrows in the $\hat{z}$ direction creates a ground-state Zeeman levels population with a nonzero second moment $\langle m^2 \rangle > 0$. The magnetic field induces coherence mixing, represented by blue arrows. The probe beam, indicated by red dashed arrows, is spatially separated from and parallel to the pump beam. It is linearly polarized and consists of $\sigma^+$ and $\sigma^-$ circular polarizations. Each polarization propagates with a different index of refraction, dependent on the Zeeman coherence. This asymmetry induces birefringence during optical resonance interactions with the atoms.

In summary, NLFE experiments depend on the interplay among three key factors: the production of spin coherence through optical pumping, the subsequent evolution of this coherence in the presence of an external magnetic field, and the optical detection of the resulting changes in the medium's properties. The specific use of linearly polarized resonance radiation in the first step allows for the preparation of $\Delta\mathrm{m} = \pm 2$ coherence. This evolves twice as fast in the interzone region as $\Delta\mathrm{m} = \pm 1$ coherence does, leading to a halving of the resonance magnetic linewidth.

As presented in [40], if the birefringence axis is oriented at an angle $\theta$, which is dependent on the B-field precession angle with respect to the linear light polarization and polarization decoherence rate, we can simplify the Faraday rotation $\Phi_F$, which depends on the birefringence of the atomic medium, as follows in (1). This expression holds under the assumptions of the small-angle approximation $(\theta \ll 1)$ , where birefringence-induced phase shifts remain linear, and the steady-state regime, ensuring that transient effects from optical pumping and atomic motion are negligible.

$$\Phi_F = \mathrm{Im}(N_{\parallel} - N_{\perp}) \frac{\omega_L L}{2c} \sin(2\theta) \tag{1}$$

Here $\mathrm{Im}(N_{\parallel} - N_{\perp})$ represent the imaginary parts of the index of refraction for the light polarized parallel, $N_{\parallel}$, and perpendicular, $N_{\perp}$, to the axis of birefringence, respectively. The separation length between the pump and the probe beam $L$, and $\omega_L$ represents the Larmor frequency. One can show that $\mathrm{Im}(N_{\parallel} - N_{\perp})$ is proportional to (2), as shown in [41]:

$$\mathrm{Im}(N_{\parallel} - N_{\perp}) \propto \langle 3F_z^2 - F^2 \rangle = \left\langle 2F_z^2 - \left(F_x^2 + F_y^2\right) \right\rangle \tag{2}$$

Eq. (2) connects the observed birefringence to the atomic alignment $\langle 3F_z^2 - F^2 \rangle$, which can be re-expressed using the angular momentum operators $\hat{F}_x$, $\hat{F}_y$, and $\hat{F}_z$. To evaluate this quantity in terms of atomic populations and coherences, we compute the macroscopic angular momentum components of the atomic ensemble. These are derived from the atomic density matrix elements $\rho_{i,j}$, which describe population and coherence between Zeeman sublevels, as shown in Eqs. (3-(5)

$$\langle \hat{F}_x \rangle = \left\langle \frac{1}{2} \left( \hat{F}_+ + \hat{F}_- \right) \right\rangle = N \sum_{m=-F}^{F-1} \frac{A(F,m)}{2} \left( \rho_{m+1,m} + \rho_{m,m+1} \right) \tag{3}$$

$$\langle \hat{F}_y \rangle = \left\langle \frac{1}{2i} \left( \hat{F}_+ - \hat{F}_- \right) \right\rangle = N \sum_{m=-F}^{F-1} \frac{A(F,m)}{2i} \left( \rho_{m+1,m} - \rho_{m,m+1} \right) \tag{4}$$

$$\langle \hat{F}_z \rangle = N \sum_{m=-F}^{F} m \rho_{m,m} \tag{5}$$

Where $A(F,m) = \sqrt{F(F+1) - m(m+1)}$, $\hat{F}_{+,-}$ are the rising and lowering operators for the spin along $\hat{z}$ respectively and $\rho_{i,j}$ is the density operator that describe the state of the atoms and $i,j = -F, \ldots, F$ parametrize the Zeeman magnetic sublevels. For more details on the density operators see [42] Part 1.

To simulate the dependence of the measured signal on magnetic field scanning, several physical phenomena must be considered. First, Zeeman splitting due to the magnetic field creates a broader structure with a linear profile around zero magnetic field, characterized by the amplitude $c_1$. Second, Ramsey fringes arise from the NLFE resonance. This behavior is modeled by the integrating over the parallel velocities $v_{\parallel}$ of the following components: the Larmor-rotating term of atomic polarization $\sin(2\omega_L t_{\perp})$; the inhomogeneous Maxwellian distribution $\mathcal{M}(v_{\parallel})$ of atomic longitudinal velocities relative to the pump beam; and various spin decoherence processes [43].

Consequently, under the assumptions of linear response, steady-state spin dynamics, absence of buffer gas, and uniform optical fields, the Faraday rotation is expressed as a velocity-averaged integral over $v_{\parallel}$, as shown in Eq. (6) [44]. The transverse propagation time $t_{\perp} = L/v_{\perp}$ is treated as a fixed parameter, constrained by the thin cell geometry, since only atoms with suitable transverse velocity components $v_{\perp}$ contribute to the signal. In contrast, the parallel velocity $v_{\parallel}$ remains broadly distributed and is integrated in the expression. The second term involves the velocity distribution, determined by cell geometry, and follows the Maxwellian profile given in Eq. (7)**.** The dimensionless parameter $\eta = L|v_{\parallel}|/Wu = \tilde{t}_{\perp}/t_{\parallel}$ represents the ratio between the average transverse and longitudinal propagation times. Here, the most probable transverse time is $\tilde{t}_{\perp} = L/u$, where $u$ is the most probable atomic speed and the longitudinal propagation time is $t_{\parallel} = W/|v_{\parallel}|$. For further explanations, see [42] part 2.

Damping of the Ramsey fringes arises from the velocity distribution and, in this case, depends on the cell width $W$ and the separation distance $L$ between the probe and pump areas. The amplitude of the Ramsey fringes is denoted $c_2$. Both amplitudes, $c_1$ and $c_2$, are functions of experimental parameters such as the optical power density, laser detuning, the number of interrogated atoms, cells properties and the specific alkali species used. The gyromagnetic ratio is defined as $\gamma = g_F \mu_B/\hbar$, where $g_F$ is the Landé g-factor and $\mu_B$ is the Bohr magnetron. Consequently, the Larmor frequency is $\omega_L = \gamma B$, where $B$ is the magnitude of the total magnetic field magnitude. Relaxation of spin alignment due to atomic collisions is modeled by the exponential decay term $\mathrm{e}^{-\Gamma t_{\perp}}$, where $\Gamma$ is the total decoherence rate. In this regime, wall collisions dominate, with $\Gamma_{\text{wall}} = \bar{v}/L_{eff}$, where $\bar{v}$ is the mean atomic velocity and $L_{eff}$ is the effective mean free path [45].

$$\Phi_{\mathrm{F}}(v_{\parallel}) \propto c_1\omega_L + c_2 \int_{-\infty}^{\infty} \mathcal{M}(v_{\parallel}) \cdot \cos(2\omega_L t_{\perp}) \cdot \mathrm{e}^{-\Gamma t_{\perp}} dv_{\parallel} \qquad (6)$$

$$\mathcal{M}(v_{\parallel}) \propto \frac{1}{\pi}\mathrm{e}^{-(\eta)^2} - \frac{1}{\sqrt{\pi}}\eta(1-\mathrm{erf}(\eta)) \qquad (7)$$

In this study, we exploit the fact that, in a thin cell, the atoms contributing to the signal predominantly move parallel to the cell walls. By applying the Faraday-Ramsey scheme, we achieve velocity filtering of propagating coherences, allowing us to measure only the atoms that maintain coherence over the cell's length. This geometry enables coherence lifetimes of $\tau = \Gamma^{-1}$, limited by wall collisions, to reach values typical of millimeter-sized cells, with $\Gamma \propto u/L$, in the order of a few microseconds. Even in ultra-thin, micron-scale cells, where $\Gamma \propto u/W$, coherence lifetimes can reach tens of nanoseconds - substantially longer than expected given the small dimensions.

## 3. Experiment

The experiments were conducted on the D2 resonance line of rubidium at $\lambda = 780\ nm$. We tuned the laser to the $F_g = 2 \rightarrow F_e = 3$ transition and stabilized it using the Doppler-free dichroic atomic vapor laser lock [46]. Fig. 3 illustrates the experimental setup.

The laser beam was split using a non-polarizing beam splitter in a 90:10 ratio. The higher-power branch served as the pump beam, which was linearly polarized and set to a power of 12 $mW$. We expanded the pump beam using a beam expander and then passed it through an axicon lens to generate a ring-shaped profile. The ring diameter could be varied by adjusting the axicon position. A linearly polarized probe beam with a diameter of $1\ mm$ and power of $150\ \mu W$, passed through the center of the pumping ring.

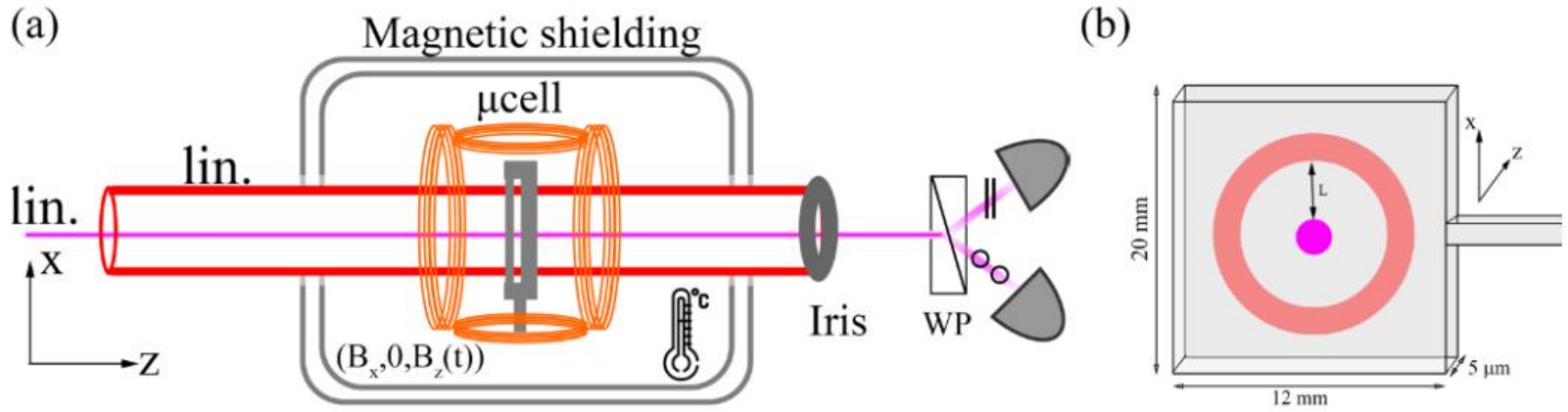

Fig. 3. (a) Cartoon of the experimental setup used for the Faraday rotation measurement. A linearly polarized pump beam (red) and a probe beam (magenta) are incidents on the vapor cell, which is heated to $120\ ℃$ and enclosed in a controlled magnetic environment created by external coils and four layers of µ-metal magnetic shielding. The pump beam is spatially filtered by an iris after passing through the cell, while the probe beam is directed to a Wollaston Prism (WP) that separates its polarization components. (b) Top-view schematic of the pump and probe illumination configuration within the cell volume.

We used a ring-shaped pump beam to exploit the cell's radial symmetry. This symmetry ensures equal travel times across the radial paths from the pump and probe regions. As a result, two key parameters stay consistent across all radial paths: the velocity distribution's dependence on the cell geometry and the Zeeman population and coherence dynamics under the applied magnetic field. By maintaining these conditions, we were able to measure a larger number of atoms with well-defined propagation characteristics, enhancing both the signal strength and the reliability of our observations.

By varying the pump beam ring diameter from $1\ mm$ to $8\ mm$, we controlled the distance between the pump and probe beams. The probe beam intensity was adjusted using variable neutral-density filters to maintain a consistent Rabi frequency ratio between the beams. In the theoretical model, we assume that atoms traverse the interaction region slowly enough to undergo spatially averaged pumping and probing. Under this assumption, we set the Rabi frequency of the pump beam to $\Omega_{pu} = 45\ MHz$ and that of the probe beam to $\Omega_{\mathrm{pr}} = 2\ MHz$ [42] Part 1. We adjusted the polarization direction to equalize the intensity of both components, while we tuned the laser's wavelength approximately $20\ MHz$ red of the transition resonance to optimize the balance between the amplitude of the polarization rotation angle and the probe's beam transparency.

We measured the Faraday rotation of the probe beam using a Wollaston prism, which separated the polarization components and collected them using two channels of a balanced photo detector. We observed a Faraday rotation signal when probe's resonant interaction with the atoms induced a Faraday rotation effect. As shown in Fig. 3, the dimensions of the homemade $^{87}$Rb vapor cell are $20\ mm$ height and $12\ mm$ width. We used a $5\ \mu m$ thick cell fabricated using lithography, reactive ion etching, and anodic bonding as described in [20]. We heated the cell up to 120 ℃, leading to a rubidium atomic density of $n_{Rb} \approx 2 \times 10^{13}\ cm^{-3}$. The estimated flux of polarized atoms at the largest pump-to-probe separation is $F_{Rb} \approx 2.2 \times 10^{9}\ sec^{-1}$ [42] Part 4.

To maintain a stable and well-defined magnetic environment, the cell was placed inside four layers of mu-metal shielding to suppress stray fields. Inside the shielding, two orthogonal Helmholtz coil pairs were used. The first generated a static magnetic field $B_{\perp} = 20\ \mu T$, perpendicular to the beam propagation axis, to mix Zeeman sublevels and lift ground-state degeneracy (Fig. 2). The second coil set generated a variable magnetic field $B_{\parallel}$ along the beam direction, which we scanned between $-250\ \mu T$ and $+250\ \mu T$. This allowed us to control the total magnetic field $B = \left(B_{\perp}^{2} + B_{\parallel}^{2}\right)^{1/2}$, thereby tuning the Larmor frequency and enabling observation of Ramsey fringes. Slow scanning of the magnetic field ($10\ Hz$), ensuring that the period of a magnetic sweep is much longer than the typical atom transit time across the cell. By measuring the average Faraday rotation as a function of the magnetic field, we obtained Faraday-Ramsey spectra, as shown in Fig. 4(a).

At $B = B_{\perp}$, the expected Larmor frequency is approximately $2\pi \times 140\ KHz$. Under experimental conditions, the atoms mean velocity is $v_p = 277\ m/sec$ ([47], part 4), meaning an atom travels roughly $1\ mm$ during one Larmor precession cycle.

## 4. Experimental results

We used wall collisions to sort atomic velocities by direction and applied specific magnetic fields to induce spin precession. As described in (6) the atoms exhibited Larmor dynamics. At an operating temperature of $120\ °C$, interatomic spin-exchange collisions [48], which lead to spin decoherence at rate $\Gamma_{\mathrm{se}}$, were not significant under our experimental conditions. These collisions contribute to the total decoherence rate $\Gamma$, which is determined by the relation $\Gamma^{-1} = \Gamma_{se}^{-1} + \Gamma_{wall}^{-1}$, where $\Gamma_{wall}$ is the decoherence rate due to wall collisions. However, during standard Larmor rotation, represented in the model by the $\cos(2\omega_L t_{\perp})$ term, atoms with different planar velocities reached the probing area at different times and because of velocities integration, relative dephasing occurred between them, and it is the dominant decoherence process in the measured conditions.

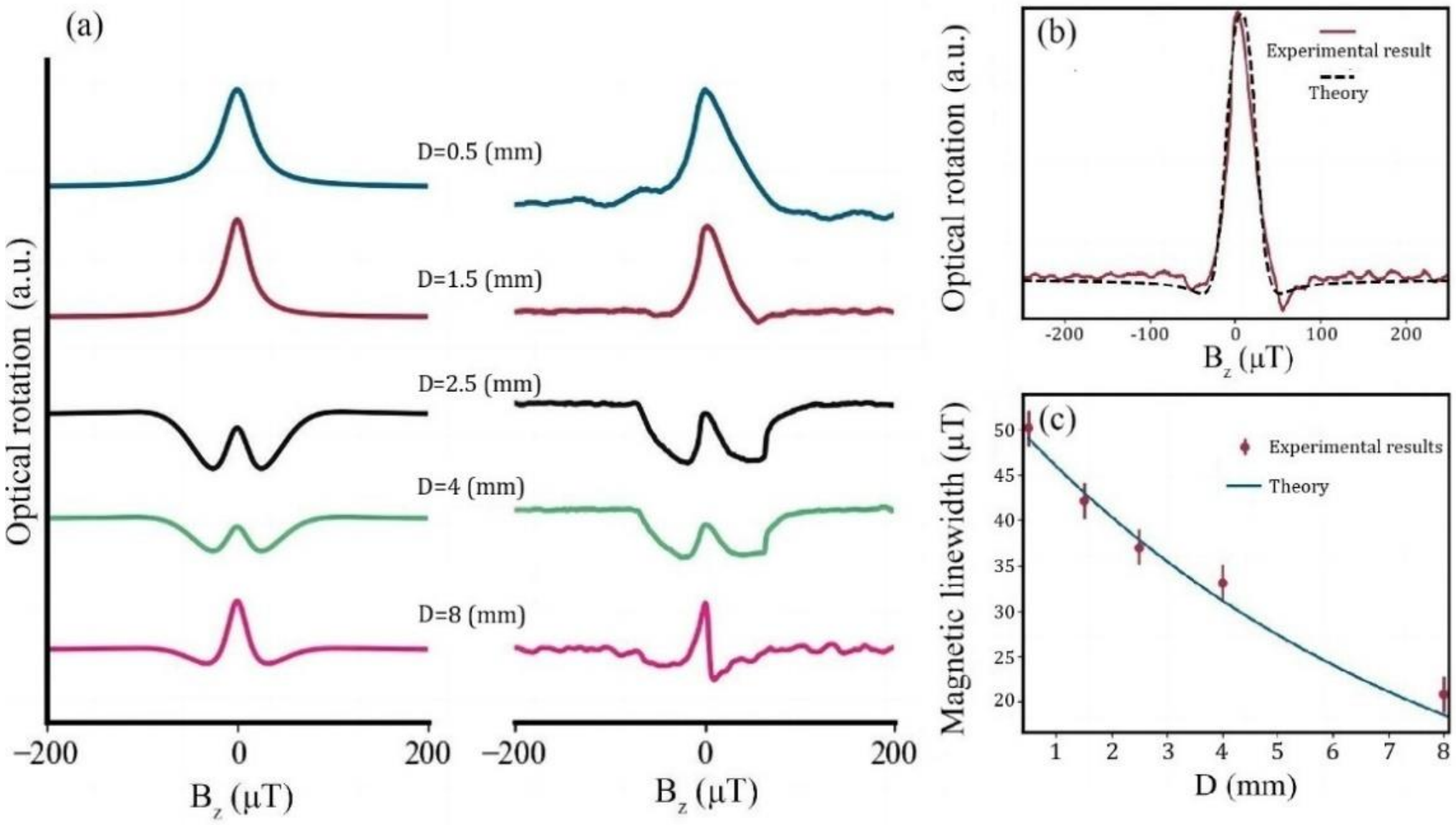

Fig. 4. (a) Theoretical (left) and experimental (right) Faraday-Ramsey rotation signals as a function of the longitudinal magnetic field $B_z$, with a constant transverse field $B_x = 20\ \mu$T. Measurements were taken for the $F_g = 2 \rightarrow F_e = 3$ transition in $^{87}$Rb with a $5\ \mu m$-thick vapor cell. Results are shown for various pump ring diameters, using linearly polarized pump and probe beams. All graphs are normalized, with offset applied for improved visualization. The smallest diameter $D = 0.5\ mm$ represents the overlapping-beam configuration without propagation. (b) Comparison of theoretical and experimental data for a pump beam diameter of 1.5 mm. The theoretical curve is normalized to match the experimental data and shows good agreement. (c) Extracted magnetic linewidths (FWHM) as a function of the pump beam inner diameter. Both measured and simulated values show a clear narrowing of the resonance as the pump-probe separation increases. We present normalized magnetic field scanning results, fitted to the corresponding model, to analyze the shape of the Ramsey-Fringes, with particular emphasis on their linewidth. The results confirm that increasing the distance between the pump and probe regions reduces the magnetic linewidth. This narrowing is attributed to enhanced velocity filtering: longer precession paths favor contributions from atoms with more uniform transverse velocities, improving coherence.

In Fig. 4(a), we present both theoretical and experimental line shapes of the Faraday-Ramsey rotation signal for a 5 $\mu m$-thick cell, plotted as a function of the $B_z$ component of the total magnetic field $B$. Here $B_z$ is scanned while maintaining a constant transverse field $B_x = 20\ \mu T$, perpendicular to the pump beam direction. This corresponds to the second term in Eq. (6) as discussed in Section 3. The first term, representing linear Faraday rotation, is removed through calibration using measurements without optical pumping, as described in [42] Part 5. In Fig. 4(b), we demonstrate strong agreement between the experimental data and theoretical predictions for a pump beam diameter of 1.5 mm and a 5 μm-thick cell.

Next, we extracted the magnetic linewidth (FWHM – Full Width Half Maximum) for all measured pump-probe separations. In Fig. 4(c) we present the results, which are consistent with both the theoretical model and the geometric intuition underlying atomic beam configurations. A potential limiting factor for at larger pump–probe separations is spin-exchange collisions between alkali atoms. Since this process depends on the vapor density, there exists an optimal separation distance approximately equal to the spin-exchange mean free path $\lambda_{SE}$ [42] Part 4. Beyond this optimal length, further increasing the separation does not significantly improve the resonance linewidth.

## 5. Conclusions

In this work we presented a compact analog to Faraday-Ramsey rotation measurements using thin alkali vapor cells fabricated via chip-scale manufacturing. These cells contain rubidium atoms optically polarized by a resonant laser beam. The thin cell geometry enables selective transmission of atoms with minimal spin decoherence due to wall collisions. Atoms propagating parallel to the cell walls behave as an atomic-beam-like ensemble, with an estimated coherent flux of $F_{Rb} \cong 2 \times 10^9\ sec^{-1}$. We observed the Faraday-Ramsey effect using a radially symmetric pumping configuration, which utilized spatial uniformity to improve the signal-to-noise ratio. The experimental results were in good agreement with theoretical predictions.

The atomic ensemble exhibits a central velocity, and a Gaussian velocity distribution due to Doppler broadening, which contributes to the magnetic resonance linewidth. In thin cells, the narrow geometry constrains the velocity distribution, enhancing the effect of spatial velocity filtering. Measurements of magnetic resonance lineshapes at different pump-probe separations fit the proposed model.

The coherent atomic flux in this thin-cell configuration is currently limited by the low alkali density, resulting in a relatively weak signal. Spin-exchange collisions among alkali atoms further constrain the maximum usable density, thereby limiting the attainable magnetic linewidth. To increase the density, we propose operating at higher temperatures and using cesium instead of rubidium. Additionally, incorporating a spatial filter, localized dispenser, and a temperature gradient may help mitigate spin-exchange limitations by reducing the density of unpolarized atoms while boosting the polarized atomic flux.

The use of thin vapor cells to emulate an atomic beam offers several advantages: it is simple, scalable, and highly compatible with device miniaturization. This approach has potential applications in fields that traditionally rely on atomic beams, such as frequency standards, magnetometry, gradiometry, and Rydberg atom research.

**Supplemental document**. See Supplemental information [42] for supporting content.

### References

1. C. S. Wood, S. C. Bennett, D. Cho, B. P. Masterson, J. L. Roberts, C. E. Tanner, and C. E. Wieman, "Measurement of Parity Nonconservation and an Anapole Moment in Cesium," Science (1979) 275(5307), 1759–1763 (1997).

2. J. P. Carrico, E. Lipworth, P. G. H. Sandars, T. S. Stein, and M. C. Weisskopf, Electric Dipole Moments of Alkali Atoms. A Limit to the Electric Dipole Moment of the Free Electron ~ (1968), 174.
3. N. F. Ramsey, "A Molecular Beam Resonance Method with Separated Oscillating Fields," Physical Review 78(6), 695–699 (1950).
4. M. Allegrini, E. Arimondo, and L. A. Orozco, "Survey of Hyperfine Structure Measurements in Alkali Atoms," J Phys Chem Ref Data 51(4), (2022).
5. E. G. Myers, "High-precision atomic mass measurements for fundamental constants," Atoms 7(1), (2019).
6. J. Glick, W. Huntington, M. Borysow, K. Wen, D. Heinzen, J. Kłos, and E. Tiesinga, "Monte Carlo simulations of the capture and cooling of alkali-metal atoms by a supersonic helium jet," Phys Rev A (Coll Park) 110(2), (2024).
7. J. Vanier, A. Godone, F. Levi, and S. Micalizio, "Atomic clocks based on coherent population trapping: basic theoretical models and frequency stability," in IEEE International Frequency Control Symposium and PDA Exhibition Jointly with the 17th European Frequency and Time Forum, 2003. Proceedings of the 2003 (2003), pp. 2–15.
8. J. D. Elgin, T. P. Heavner, J. Kitching, E. A. Donley, J. Denney, and E. A. Salim, "A cold-atom beam clock based on coherent population trapping," Appl Phys Lett 115(3), 033503 (2019).
9. M. A. Lombardi, T. P. Heavner, and S. R. Jefferts, "NIST Primary Frequency Standards and the Realization of the SI Second," NCSLI Measure 2(4), 74–89 (2007).
10. C. Carlé, M. A. Hafiz, S. Keshavarzi, R. Vicarini, N. Passilly, and R. Boudot, "Pulsed-CPT Cs-Ne microcell atomic clock with frequency stability below 2 × 10−12 at 105 s," Opt. Express 31(5), 8160–8169 (2023).
11. S. Knappe, V. Gerginov, P. D. D. Schwindt, V. Shah, H. G. Robinson, L. Hollberg, and J. Kitching, Atomic Vapor Cells for Chip-Scale Atomic Clocks with Improved Long-Term Frequency Stability (2005).
12. K. W. Martin, G. Phelps, N. D. Lemke, M. S. Bigelow, B. Stuhl, M. Wojcik, M. Holt, I. Coddington, M. W. Bishop, and J. H. Burke, "Compact Optical Atomic Clock Based on a Two-Photon Transition in Rubidium," Phys Rev Appl 9(1), (2018).
13. A. Fabricant, I. Novikova, and G. Bison, "How to build a magnetometer with thermal atomic vapor: a tutorial," New J Phys 25(2), 025001 (2023).
14. O. Alem, R. Mhaskar, R. Jiménez-Martínez, D. Sheng, J. LeBlanc, L. Trahms, T. Sander, J. Kitching, and S. Knappe, "Magnetic field imaging with microfabricated optically-pumped magnetometers," Opt Express 25(7), 7849 (2017).
15. R. M. Noor and A. M. Shkel, "MEMS Components for NMR Atomic Sensors," Journal of Microelectromechanical Systems 27(6), 1148–1159 (2018).
16. T. W. Kornack, R. K. Ghosh, and M. V. Romalis, "Nuclear spin gyroscope based on an atomic comagnetometer," Phys Rev Lett 95(23), (2005).
17. J. Fang, S. Wan, and H. Yuan, "Dynamics of an all-optical atomic spin gyroscope," Appl Opt 52(30), 7220–7227 (2013).
18. Y. Gao, D. Ma, S. Jiang, Y. Ma, S. Li, K. Wang, and X. Xu, "A novel oven structure for improving temperature uniformity of vapor cell in atomic sensors," Results Phys 47, 106339 (2023).
19. J. Kitching, "Chip-scale atomic devices," Appl Phys Rev 5(3), 031302 (2018).
20. E. Talker, R. Zektzer, Y. Barash, N. Mazurski, and U. Levy, "Atomic spectroscopy and laser frequency stabilization with scalable micrometer and sub-micrometer vapor cells," Journal of Vacuum Science & Technology B, Nanotechnology and Microelectronics: Materials, Processing, Measurement, and Phenomena 38(5), (2020).
21. E. Talker, P. Arora, M. Dikopoltsev, and U. Levy, "Optical isolator based on highly efficient optical pumping of Rb atoms in a miniaturized vapor cell," Journal of Physics B: Atomic, Molecular and Optical Physics 53(4), 045201 (2020).
22. Y. Sebbag, Y. Barash, and U. Levy, "Generation of coherent mid-IR light by parametric four-wave mixing in alkali vapor," in 2018 Conference on Lasers and Electro-Optics (CLEO) (2018), pp. 1–2.
23. D. Sheng, A. R. Perry, S. P. Krzyzewski, S. Geller, J. Kitching, and S. Knappe, "A microfabricated optically-pumped magnetic gradiometer," Appl Phys Lett 110(3), (2017).
24. J. A. Gordon, C. L. Holloway, A. Schwarzkopf, D. A. Anderson, S. Miller, N. Thaicharoen, and G. Raithel, "Millimeter wave detection via Autler-Townes splitting in rubidium Rydberg atoms," Appl Phys Lett 105(2), (2014).
25. G. D. Martinez, C. Li, A. Staron, J. Kitching, C. Raman, and W. R. McGehee, "A chip-scale atomic beam clock," Nat Commun 14(1), 3501 (2023).
26. S. Briaudeau, D. Bloch, and M. Ducloy, Sub-Doppler Spectroscopy in a Thin Film of Resonant Vapor (n.d.).
27. S. Briaudeau, D. Bloch, and M. Ducloy, "Detection of slow atoms in laser spectroscopy of a thin vapor film," Europhysics Letters (EPL) 35(5), 337–342 (1996).
28. A. Sargsyan, A. Tonoyan, R. Mirzoyan, D. Sarkisyan, A. M. Wojciechowski, A. Stabrawa, and W. Gawlik, "Saturated-absorption spectroscopy revisited: atomic transitions in strong magnetic fields (>20 mT) with a micrometer-thin cell," Opt Lett 39(8), 2270 (2014).

29. A. Sargsyan, D. Sarkisyan, and A. Papoyan, "Dark-line atomic resonances in a submicron-thin Rb vapor layer," Phys Rev A 73(3), (2006).
30. R. Boudot, D. Miletic, P. Dziuban, C. Affolderbach, P. Knapkiewicz, J. Dziuban, G. Mileti, V. Girodano, and C. Gorecki, "First-order cancellation of the Cs clock frequency temperature-dependence in Ne-Ar buffer gas mixture," Opt Express 19(4), 3106 (2011).
31. L. Wu, J. Shang, Y. Ji, Q. Gan, and C.-P. Wong, "Influence of Buffer-Gas Pressure Inside Micro Alkali Vapor Cells on the Performance of Chip-Scale SERF Magnetometers," IEEE Trans Compon Packaging Manuf Technol 8(4), 621–625 (2018).
32. A. Berrebi, M. Dikopoltsev, O. Katz, and O. Katz, "Optical protection of alkali-metal atoms from spin relaxation," ArXiv (2022).
33. O. Katz, M. Dikopoltsev, O. Peleg, M. Shuker, J. Steinhauer, and N. Katz, "Nonlinear elimination of spin-exchange relaxation of high magnetic moments," Phys Rev Lett 110(26), (2013).
34. S. J. Seltzer, D. J. Michalak, M. H. Donaldson, M. V. Balabas, S. K. Barber, S. L. Bernasek, M.-A. Bouchiat, A. Hexemer, A. M. Hibberd, D. F. J. Kimball, C. Jaye, T. Karaulanov, F. A. Narducci, S. A. Rangwala, H. G. Robinson, A. K. Shmakov, D. L. Voronov, V. V. Yashchuk, A. Pines, and D. Budker, "Investigation of antirelaxation coatings for alkali-metal vapor cells using surface science techniques," J Chem Phys 133(14), (2010).
35. R. Straessle, M. Pellaton, C. Affolderbach, Y. Pétremand, D. Briand, G. Mileti, and N. F. de Rooij, "Microfabricated alkali vapor cell with anti-relaxation wall coating," Appl Phys Lett 105(4), (2014).
36. J. Lazare, N. Enakaya, and J. L. Lyons, "Evaluation of alternatives to paraffin for antirelaxation coatings," Journal of Vacuum Science & Technology A: Vacuum, Surfaces, and Films 38(3), (2020).
37. D. Budker, W. Gawlik, D. F. Kimball, S. M. Rochester, V. V. Yashchuk, and A. Weis, "Resonant nonlinear magneto-optical effects in atoms," Rev Mod Phys 74(4), 1153–1201 (2002).
38. Alexandrov Evgeniy B., Auzinsh Maris, Budker Dmitry, Kimball Derek F., Rochester Scott M., and Yashchuk Vladimir V., "Dynamic effects in nonlinear magneto-optics of atoms and molecules: review," Journal of the Optical Society of America B 22(1), 7–20 (2005).
39. B. Schuh, S. I. Kanorsky, A. Weis, and T. W. Hänsch, "Observation of Ramsey fringes in nonlinear Faraday rotation," Opt Commun 100(5–6), 451–455 (1993).
40. A. Weis, J. Wurster, and S. I. Kanorsky, "Quantitative interpretation of the nonlinear Faraday effect as a Hanle effect of a light-induced birefringence," Journal of the Optical Society of America B 10(4), 716–724 (1993).
41. S. I. Kanorsky, A. Weis, J. Wurster, and T. W. Hänsch, "Quantitative investigation of the resonant nonlinear Faraday effect under conditions of optical hyperfine pumping," Phys Rev A (Coll Park) 47(2), 1220–1226 (1993).
42. Dikopoltsev Mark, Talker Eliran, Barash Yefim, Mazurski Noa, and Levy Uriel, "Supplementary information - Faraday-Ramsey rotation measurement in a thin cell as analogy to atomic beam," (2024).
43. J. C. Allred, R. N. Lyman, T. W. Kornack, and M. V. Romalis, "High-sensitivity atomic magnetometer unaffected by spin-exchange relaxation," Phys Rev Lett 89(13), 1308011–1308014 (2002).
44. S. Nakayama, G. W. Series, and W. Gawlik, "Larmor precession in polarization spectroscoy with satially separated beams," Opt Commun 34(3), 389–392 (1980).
45. N. Castagna, G. Bison, G. Di Domenico, A. Hofer, P. Knowles, C. MacChione, H. Saudan, and A. Weis, "A large sample study of spin relaxation and magnetometric sensitivity of paraffin-coated Cs vapor cells," Appl Phys B 96(4), 763–772 (2009).
46. K. L. Corwin, Z.-T. Lu, C. F. Hand, R. J. Epstein, and C. E. Wieman, "Frequency-stabilized diode laser with the Zeeman shift in an atomic vapor," Appl Opt 37(15), 3295 (1998).
47. M. Dikopoltsev, E. Talker, Y. Barash, N. Mazurski, and U. Levy, Supplementary Information Faraday-Ramsey Rotation Measurement in a Thin Cell as Analogy to Atomic Beam A. Optical Bloch Equation (OBE) (2024).
48. O. Katz, O. Peleg, and O. Firstenberg, "Coherent Coupling of Alkali Atoms by Random Collisions," Phys Rev Lett 115(11), (2015).

## FARADAY-RAMSEY ROTATION MEASUREMENT IN A THIN CELL AS AN ANALOGY TO AN ATOMIC BEAM: SUPPLEMENTAL DOCUMENT

### 1. Optical Bloch Equation (OBE)

We consider atoms illuminated by laser light resonant with the $F_g = 2 \rightarrow F_e = 3$ transition on the $D_2$ line of $^{87}$Rb. A static magnetic field $B_z$ is applied collinearly with the laser propagation direction to define the quantization axis ($B_\parallel$) and a perpendicular magnetic field $B_x$ is simultaneously applied ($B_\perp$). A linearly polarized probe beam propagates along the $B_\parallel$ axis.

We describe the atomic system using the master equation for the density matrix:

$$\partial_t \rho = -\frac{i}{\hbar}[\mathcal{H}_B + \mathcal{H}_0 + \mathcal{V}(t), \rho] + \mathcal{L}\{\rho\} \tag{S1}$$

The unperturbed Hamiltonian, diagonal in the Zeeman basis, is:

$$\mathcal{H}_0 = E_0(F) \tag{S2}$$

Which corresponds to the energy of the atomic sublevels. The Zeeman interaction Hamiltonian is:

$$\mathcal{H}_B = g_F \mu_B \vec{F} \cdot \vec{B} \tag{S3}$$

In our configuration, a slowly varying magnetic field $B_z(t)\hat{z}$ applied along with a transverse magnetic field $B_x$ to induce Zeeman coherence mixing. This gives rise to the single-atom Hamiltonian:

$$\mathcal{H}_B = g_F \mu_B (B_z F_z + B_x F_x) \tag{S4}$$

The dipole interaction with light, under the rotating-wave approximation, is described by:

$$\mathcal{V}(t) = \frac{\hbar}{2\sqrt{2}}\left[\sum_{g_k} \Omega_{pu}(t)\left(\rho_{g_k, g_{k\pm1}}\right) + \Omega_{pr}(t)\left(\rho_{g_k, g_{k\pm1}}\right)\right] + C.C. \tag{S5}$$

The Rabi frequencies of the pump and probe beams are:

$$\Omega_{pu}(t) = \Omega_{pu} \cdot rect\left(\frac{t - T_1}{\frac{T_1}{2}}\right)\exp\left(-i\omega_{pu}t\right) + c.c. \tag{S6}$$

$$\Omega_{pr}(t) = \Omega_{pr} \cdot rect\left(\frac{t - (T_2 + \frac{3}{2}T_1)}{\frac{T_1}{2}}\right) exp\left(-i\omega_{pr}t\right) + c.c. \tag{S7}$$

This interaction Hamiltonian is the microscopic origin of the spin coherence responsible for the nonlinear Faraday rotation signal described in equation (1) and equation (6) of the main text. We use a single laser, split unequally in power, with $\Omega_{pr} < \Omega_{pu}$ and match frequencies $\omega_{pu} = \omega_{pr}$. Here, $T_1$ is the atomic time-of-flight across the beam with a width of 0.5 mm, and $T_2$ is the "dark" interval during which the intensities of both the pump and the probe beams are zero. The Liouville operator $\mathcal{L}$ describes relaxation processes in the system.

We can now express the OBE for the multilevel Zeeman system using the basis states $|e_i\rangle = |F_e, m_i\rangle, |g_i\rangle = |F_g, m_i\rangle$, as:

$$\left.\dot{\rho}_{e_i e_j}\right|_{B_x} = -i\frac{\mu_B B_x}{2\hbar}\, g_e\{c\} \tag{S8}$$

$$\dot{\rho}_{e_i,e_j} = -\left(i\omega_{e_i,e_j} + \Gamma\right)\rho_{e_i,e_j} + \frac{i}{\hbar}\sum_{g_k}\left(\rho_{e_i g_k}V_{g_k e_j} - V_{e_i g_k}\rho_{g_k e_j}\right) \tag{S9}$$

$$\dot{\rho}_{e_i g_j} = -\left(i\omega_{e_i g_j} + \frac{\Gamma}{2}\right)\rho_{e_i g_j} + \frac{i}{\hbar}\left(\sum_{e_k}\rho_{e_i e_k}V_{e_k g_j} - \sum_{g_k}V_{e_i g_k}\rho_{g_k g_j}\right) \tag{S10}$$

$$\dot{\rho}_{g_i g_j} = -i\omega_{g_i,g_j}\rho_{g_i,g_j} + \frac{i}{\hbar}\sum_{e_k}\left(\rho_{g_i e_k}V_{e_k g_j} - V_{g_i e_k}\rho_{e_k g_j}\right) + \left(\frac{d}{dt}\rho_{g_i g_j}\right)_{SE} \tag{S11}$$

Spontaneous emission repopulation of the ground state is described by:

$$\left(\frac{d}{dt}\rho_{g_i g_j}\right)_{SE} = (2F_e + 1)\Gamma \sum_{\substack{(q,q'=-F_e,F_e),\\ (p=-1,1)}} (-1)^{p-k-q'}\begin{pmatrix} F_g & 1 & F_e \\ -k & p & q \end{pmatrix}\rho_{e_q e_{q'}}\begin{pmatrix} F_e & 1 & F_g \\ -q' & -p & k' \end{pmatrix} \tag{S12}$$

By solving the above equation, we compute the expectation values of the macroscopic angular momentum components $\langle\hat{F}_x\rangle$, $\langle\hat{F}_y\rangle$, and $\langle\hat{F}_z\rangle$ in the hyperfine ground state. These components relate atomic polarization to the observed birefringence and Faraday rotation, as described in equations (3)–(5) of the main text.

## 2. Transverse velocity distribution

All calculations and plots in this document correspond to measurements performed using vapor cell with a thickness of $5\mu m$. Here, we present a simplified two-dimensional model for the transverse velocity distribution, which reflects the radial symmetry of our experimental configuration.

The longitudinal and transverse velocities of the atoms are correlated due to their direction of motion (see Fig. 1 in the main text)

$$\frac{v_\parallel}{v_\perp} = \tan(\alpha) \tag{S13}$$

For atoms that contribute to the signal, the condition $v_\parallel \ll v_\perp$ holds, since the cell width $W$ is much smaller than its length $L$, as dictated by the cell geometry. Consequently, there exists a maximum angle $\alpha_{max}$ defined by:

$$\tan(\alpha_{max}) = \frac{W}{L} = \frac{v_\parallel^{max}}{v_\perp^{min}} \rightarrow v_{\perp,min} = \frac{L}{W}v_\parallel^{max} \tag{S14}$$

Using the small-angle approximation, the transverse velocity element can be written as $dv_\perp \approx v d\alpha dv$. The distribution of atoms with transverse velocity $v_\perp$ and longitudinal velocity $v_\parallel$ is given by:

$$f(v_\parallel, v_\perp)dv_\parallel dv_\perp = f(\alpha)f_{MB}(v)d\alpha dv dv_\parallel \tag{S15}$$

Here, $f(\alpha)$ is the angular distribution function (approximated later in (S20)), and $f_{MB}(v)$ is the Maxwell-Boltzmann velocity distribution:

$$f_{MB}(v) = \pi^{-\frac{1}{2}} \cdot \frac{v}{u^2} \exp\left(-\frac{v^2}{u^2}\right) \quad \text{(S16)}$$

where $u = \sqrt{k_B T/m}$ is the most probable atomic velocity, $k_B$ is the Boltzmann constant, $T$ is the atomic temperature, and $m$ is the atomic mass.

The transverse velocity distribution projected onto $v_\parallel$ is then:

$$f_\parallel dv_\parallel = \int_{v_{\perp,min}}^{\infty} A\left(W - \frac{L \cdot v_\parallel}{v}\right) \pi^{-\frac{1}{2}} \cdot \frac{v}{u^2} exp\left(-\frac{v^2}{u^2}\right) dv dv_\parallel \quad \text{(S17)}$$

Performing the integration over $dv$ in (S17), we obtain the differential form of the longitudinal velocity distribution:

$$\mathcal{M}(v_\parallel) dv_\parallel = f_\parallel dv_\parallel \approx \frac{\sqrt{\pi} AW}{2} \left(\frac{1}{\pi} \mathrm{e}^{-(\eta)^2} - \frac{1}{\sqrt{\pi}} \eta (1 - \mathrm{erf}(\eta))\right) dv_\parallel \quad \text{(S18)}$$

Here, $\mathcal{M}(v_\parallel)$ represents the probability density function, and $dv_\parallel$ is the differential element. The dimensionless parameter $\eta = L|v_\parallel|/Wu = \tilde{t}_\perp/t_\parallel$ reflects the ratio between the transverse and longitudinal propagation times: $\tilde{t}_\perp = L/u$ the average transverse time; $t_\parallel = W/|v_\parallel|$ the longitudinal time. This velocity distribution function $\mathcal{M}(v_\parallel)$ corresponds to the one used in equation (6) of the main text to compute the velocity-averaged Faraday rotation signa and forms the basis for equation (7) in the main text.

In Fig. S1, we plot the transverse velocity distribution for 5 $\mu m$ thick cell. The transverse velocity is typically bellow $10\, m/sec$ indicating that the atoms contributing to the signal propagating nearly parallel to the cell walls — a grazing-incidence geometry favorable maintaining spin coherence.

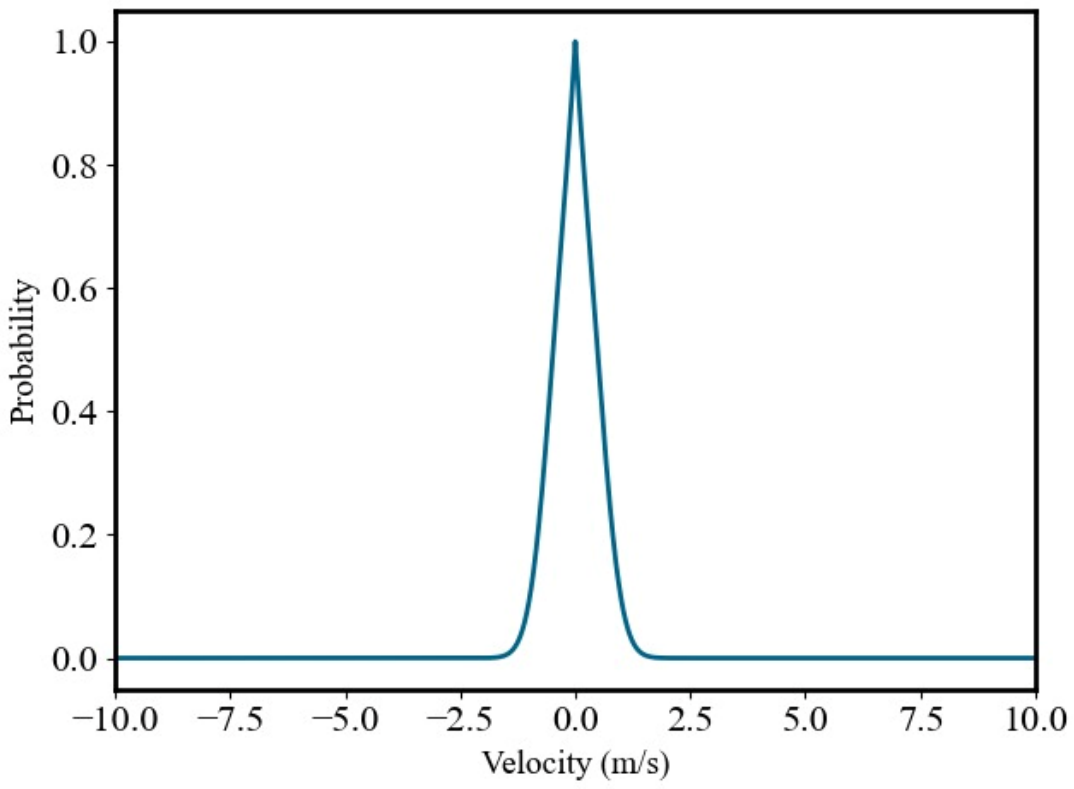

Fig. S1. Simulated transverse velocity distribution of atoms in a $5\mu m$-thick vapor cell. The distribution is sharply peaked near zero velocity, indicating that only atoms with grazing-incidence trajectories ($\lesssim 1.5\, m/sec$) contribute significantly to the signal. This reflects strong velocity filtering imposed by the thin-cell geometry, where atoms with larger transverse velocities are suppressed due to wall collisions.

### 3. Angular distribution

Consider the two-dimensional schematic shown Fig. S2, illustrating the propagation geometry for a specific angle $\alpha = \arctan\left(\frac{W-z}{L}\right)$, where $z = W - L \cdot tan(\alpha)$. Only atoms originating from the segment $l_s = z$ can reach the probe region along this trajectory.

Assuming a uniform particle density throughout the cell, the number of atoms within a small angular interval $d\alpha$ is proportional to:

$$f(\alpha)d\alpha = A\big(W - L \cdot \tan(\alpha)\big)d\alpha$$
$$-\arctan\left(\frac{W}{L}\right) \leq \alpha \leq \arctan\left(\frac{W}{L}\right) \tag{S19}$$

Here, $A$ is a normalization constant. We extended the range of $\alpha$ to include negative angles to account for symmetric propagation paths on both sides of the probe. For a thin cell ($W \ll L$), the distribution function can be linearized using the small-angle approximation:

$$f(\alpha)d\alpha \approx A(W - L \cdot \alpha)d\alpha \tag{S20}$$

To normalize the distribution, we define $N_{pu}$ as the total number of atoms pumped. Integrating over the full angular range gives:

$$N_{pu} = 2\int_0^{W/L} f(\alpha)d\alpha = A\frac{W^2}{L} \tag{S21}$$

The total pumped volume $V_{pu}$ in the 5 $\mu m$-thick cell is determined by the geometry of the ring-shaped pump beam, with width $w_{pu} = 0.5\ mm$ and inner diameter $D = 8\ mm$. Since the probe is centered at radius $L = D/2 = 4\ mm$, the pumped volume is:

$$V_{pu} = 2\pi L w_{pu} W = 6.3 \times 10^{-5}\ cm^3 \tag{S22}$$

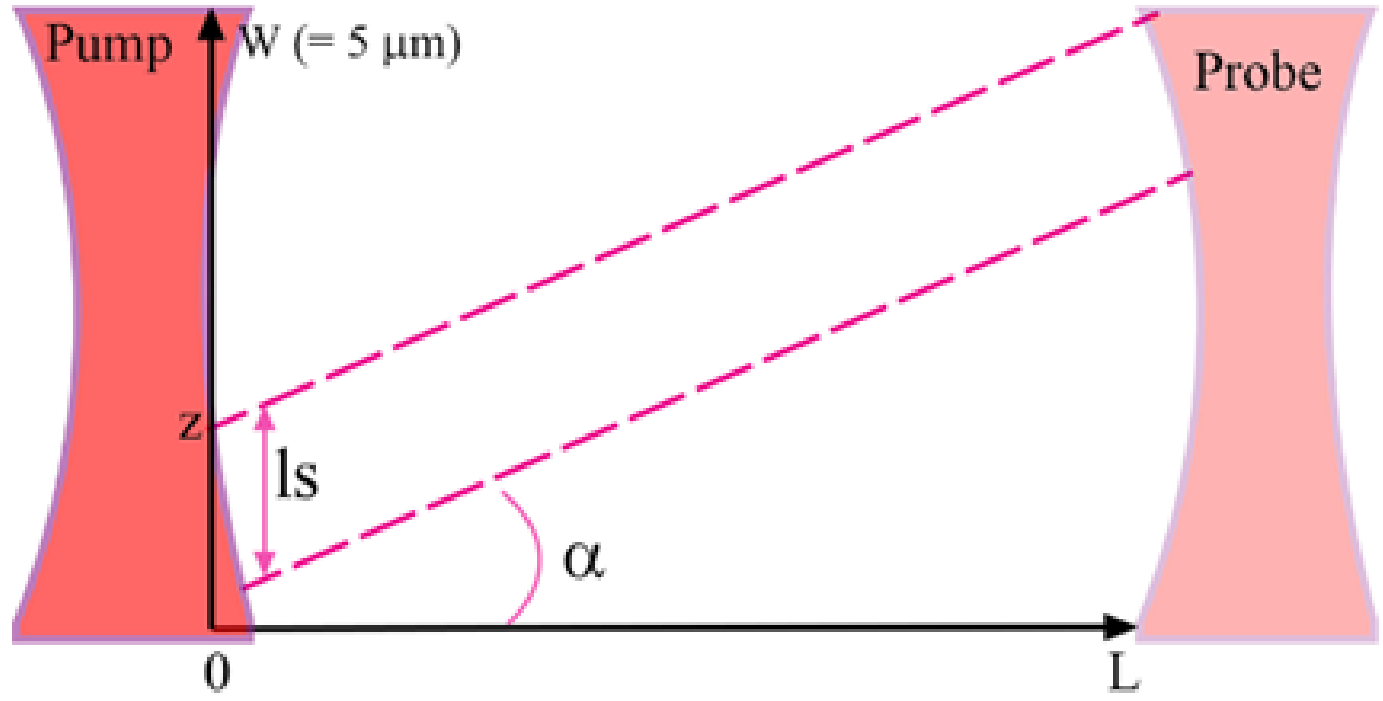

Fig. S2 Schematic side view of the pump and probe beam configuration. For a given angle $\alpha \approx (W - z)/L$, only atoms originating from the segment $l_s = z$ of the pumped region can reach the probe beam. The diagram illustrates the angular constraint imposed by the thin cell geometry, which contributes to spatial velocity filtering.

Multiplying this volume by the rubidium vapor density at the working point $n_{Rb}(120\ ^{o}C) \approx 2 \times 10^{13}\ cm^{-3}$, we obtain the total number of pumped atoms:

$$N_{pu} = n_{Rb} \cdot V_{pu} \approx 1.3 \times 10^{9} \tag{S23}$$

### 4. Coherent Atomic Flux Estimation

The number of pumped atoms, given by (S23) $N_{pu} = n_{Rb} \cdot V_{pu}$. To estimate the number of atoms that successfully reach the probe beam, we use the geometrical approximation illustrated in Fig. S3. Assuming isotropic velocity directions within the pumped volume, we model atomic propagation using spherical geometry. For each pumped atom, the possible propagation directions from a hemisphere of radius $L$. Where $L$ is the distance between the pump and probe regions.

The total spherical surface area is:

$$S_{sphere} = 4\pi L^{2} \tag{S24}$$

The effective cross-sectional area of the probe beam (as seen from the pumped atom) is approximated by:

$$S_{probe-side} = 2R_{PR} \cdot W \tag{S25}$$

where $R_{PR}$ is the probe beam radius and $W$ is the cell thickness. The probability that a pumped atom reaches the probe beam without wall collisions is given by the area ratio:

$$R_{es} = \frac{R_{PR}W}{2\pi L^{2}} \tag{S26}$$

Hence, the number of atoms expected to coherently reach the probe is:

$$N_{exp} = R_{es} \cdot N_{pu} \tag{S27}$$

To estimate the atomic flux, we assume the atoms propagate with the most probable velocity:

$$v_{p} = \sqrt{\frac{2k_{b}T}{\mathrm{m}_{rb}}} = 277\ m/sec \tag{S28}$$

The resulting atomic flux is then:

$$F_{meas} = \frac{N_{exp} \cdot v_{p}}{L} \tag{S29}$$

Using a 5 $\mu m$-thick cell, a 0.5 $mm$ pump beam width and a 0.5 $mm$ probe beam radius, we estimate the atomic flux as:

$$F_{meas} = 2.2 \times 10^{9}\ \text{atoms}/sec \tag{S30}$$

This value represents the coherent atomic flux contributing to the Faraday-Ramsey signal and complements the signal model presented in equation (6) of the main text.

In alkali vapor cells with high atomic density, the dominant decoherence mechanism is typically spin-exchange collisions, which conserve total angular momentum [1]. The mean free path for spin exchange interaction is given by:

$$\lambda_{\mathrm{SE}} = \frac{1}{\sqrt{2} \cdot n_{Rb} \cdot \sigma_{SE}} \tag{S31}$$

where $\sigma_{\mathrm{SE}} = 1.9 \times 10^{-14}\, cm^2$ is the cross-section for spin-exchange collisions. The corresponding spin-exchange rate is:

$$\mathrm{R}_{\mathrm{SE}} = \frac{V_{Rb}}{\lambda_{\mathrm{SE}}} \tag{S32}$$

where the relative thermal velocity of the rubidium atoms:

$$V_{Rb} = \sqrt{\frac{4k_b T}{\pi \mathrm{m}_{rb}}} \tag{S33}$$

with rubidium atom's mass - $\mathrm{m}_{rb}$. At the working temperature of $120\,^{o}C$, the mean free path and rate are:

$$\lambda_{\mathrm{SE}}(120\,^{o}C) \approx 18\, mm,\ \ \mathrm{R}_{\mathrm{SE}}(120^{o}C) \approx 1.2 \times 10^{4}\, sec^{-1} \tag{S34}$$

These values help determine the limits of coherence in the presence of spin-exchange interactions.

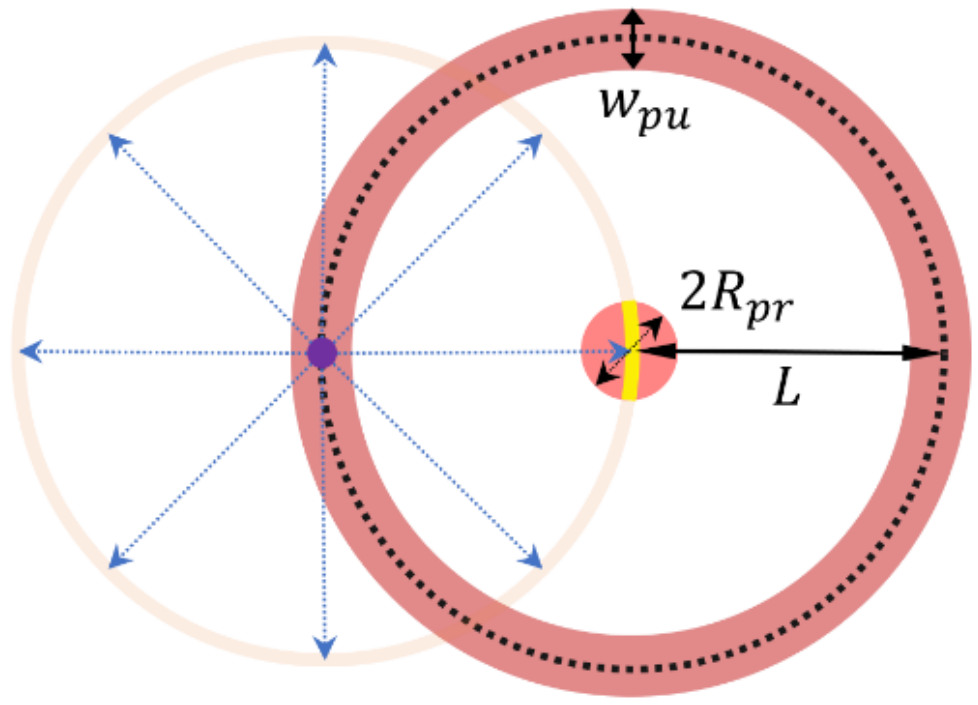

Fig. S3 . Schematic 2D sketch illustrating atomic propagation from the pumped region. Isotropic velocity directions are assumed. Here $L$ denotes the radius of the pump ring, $w_{pu}$ is the pump beam width, and $R_{pr}$ is the probe beam radius. The purple dot represents a single rubidium atom, while the blue dotted arrows indicate possible propagation directions, when part of them are towards the probe.

## 5. Nonlinear signal isolation

The Faraday effect describes the rotation of the polarization plane of linearly polarized light as it propagates through a medium subjected to a magnetic field. In hot alkali vapors, both linear and nonlinear components of the faraday effect can be observed, depending on atomic polarization, light intensity, and transition width [2]. In our work, these two components are incorporated into the model described in equation (6) of the main text.

To isolate the nonlinear component of the Faraday rotation signal, we measured the response under two experimental conditions, as shown in Fig. S4. In the first condition, the pumping beam was turned off, so no spin polarization was generated. As a result, the observed signal arose solely from the linear faraday effect, which originates from Zeeman splitting under the applied magnetic field. This measurement serves as a baseline for the system's linear optical response.

In the second condition, the pumping beam was turned on, creating spin-polarized atoms that propagated coherently across the cell. These atoms were filtered by the cell geometry and reached the probe region without significant decoherence from wall collisions. The resulting signal includes an additional nonlinear Faraday component, originating from the induced atomic alignment and coherence. By subtracting the "pump-off" trace from the "pump-on" trace, we extract the pure nonlinear Faraday rotation, isolating the coherent contribution to the signal.

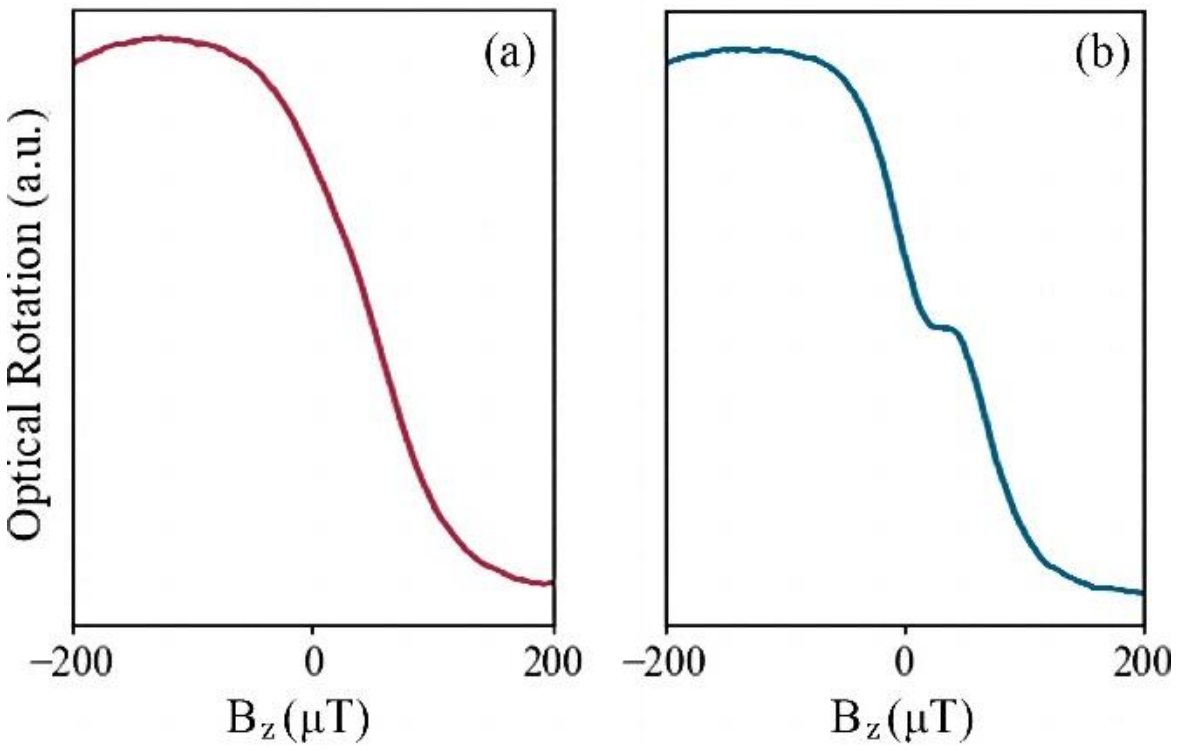

Fig. S4. Measured Faraday rotation spectra as a function of the longitudinal magnetic field $B_z$, with a constant transverse field $B_x = 20\ \mu T$, under two conditions: (a) Pumping beam off – only the linear Faraday effect is observed as the magnetic field is scanned; (b) Pumping beam on – spin-polarized atoms propagate coherently into the probe region, producing an additional nonlinear contribution. The difference between two traces reveals the isolated nonlinear Faraday rotation signal. Both spectra are normalized to the 'Pump off' responce.

## References

1. N. W. Ressler, R. H. Sands, and T. E. Stark, *Measurement of Spin-Exchange Cross Sections for Cs133, Rb87, Rb85, K39, and Na*, Physical Review **184**, 102 (1969).
2. D. Budker and M. Romalis, Optical *Magnetometry*, Nat Phys **3**, 227 (2007).